\documentclass[twocolumn,showpacs,amsmath,aps]{revtex4}

\usepackage{graphicx}
\usepackage{dcolumn}
\usepackage{bm}

\begin{document}

\title{Efficient magneto-optical trapping of Yb atoms with a violet laser diode}

\author{Chang Yong Park and Tai Hyun Yoon}
\email{thyoon@kriss.re.kr}
\affiliation{Center for Optical Frequency Control\\ Korea Research Institute of Standards and Sciecne\\
1 Doryong, Yuseong, Daejeon 305-340, Korea}
\date{July 31, 2003}

\begin{abstract}
We report the first efficient trapping of rare-earth Yb atoms with a high-power violet laser diode
(LD). An injection-locked violet LD with a 25 mW frequency-stabilized output was used for the
magneto-optical trapping (MOT) of fermionic as well as bosonic Yb isotopes. A typical number of
$4\times 10^6$ atoms for $^{174}$Yb with a trap density of $\sim 1\times10^8/$cm$^3$ was obtained.
A 10 mW violet external-cavity LD (ECLD) was used for the one-dimensional (1D) slowing of an
effusive Yb atomic beam without a Zeeman slower resulting in a 35-fold increase in the number of
trapped atoms. The overall characteristics of our compact violet MOT, e.g., the loss time of 1 s,
the loading time of 400 ms, and the cloud temperature of 0.7 mK, are comparable to those in
previously reported violet Yb MOTs, yet with a greatly reduced cost and complexity of the
experiment.
\end{abstract}

\maketitle

Atomic rare-earth ytterbium (Yb) has been proposed as one of the ideal neutral atomic systems for a
future optical frequency standard \cite{Hall} due to its unique energy level structure that is very
similar to the alkali-earth counterparts, Ca \cite{Binnewies,Oates}, Sr \cite{Xu}, and Mg
\cite{Ruschewitz}, which have a strong closed dipole-allowed transition for laser manipulation
along with a narrow intercombination transition for optical clock excitation. Recently, Katori
proposed a unique approach to develop an optical frequency standard based on the $^1$S$_0-^3$P$_0$
doubly forbidden intercombination transition in the $^{87}$Sr fermionic isotope \cite{Katori}. In
the proposal, cold fermions are trapped within the Lamb-Dicke region in a Stark shift-free optical
trap \cite{Cho}, which is believed to be an ideal cold atomic system for a long interrogation time
without residual Doppler shift, thus providing a low systematic uncertainty comparable to the
single-ion-based optical frequency standard \cite{Diddams}. To realize this proposal, a
recoil-limited laser cooling of $^{87}$Sr atoms near the Fermi temperature in an optical lattice
and a recoil-free spectroscopy of neutral Sr atoms in the Lamb-dicke regime have been demonstrated
\cite{Mukaiyama,Ido}. Quite recently, Courtillot et. al. successfully detected the weak
$^1$S$_0-^3$P$_0$ transition in a conventional $^{87}$Sr MOT and measured the absolute frequency of
the transition with a femtosecond comb \cite{Court}.

In the Yb atomic system, there are two stable fermionic isotopes, i.e., $^{171}$Yb ($I$ = 1/2) and
$^{173}$Yb ($I$ = 5/2), where $I$ is the nuclear spin. We believe that for the fermionic Yb
isotopes, Katori's proposal could equally be applicable for both fermions and even $^{171}$Yb may
compete with $^{87}$Sr for the future optical frequency standard due to its simple magnetic
sub-level structure. For this application, efficient 1D slowing and trapping of Yb atoms using a
resonant $^1$S$_0-^1$P$_1$ transition at 398.9 nm can be the starting point of the whole road to an
Yb optical clock. Recently, high-density trapping and the first Bose-Einstein condensation of Yb
atoms by an optical dipole trap loaded from an intercombination MOT (555.8 nm)
\cite{Takasu01,Takasu02} and sub-Doppler cooling effects of the Fermionic Yb atoms \cite{Maru} have
been reported. We can expect other interesting applications using the trapped Yb atoms, if one can
build a compact violet laser system at 399 nm which can be used for the efficient 1D slowing and
first-stage violet MOT \cite{Honda,Loftus,Rapol} as demonstrated in this study. Usually, a
frequency-doubled Ti:Sapphire laser is used for that purpose, but the system is complex and
expensive; therefore a study of a compact yet inexpensive violet LD system for the efficient
trapping of Yb atoms would be of importance for future applications.

With the recent developments of violet LD technologies, a high-power (30 mW) violet LD is now
commercially available. Based on this high-power violet LD, we have reported previously a
frequency-stabilized, high-power, violet LD system for the manipulation of Yb atoms
\cite{Kim,Park}. In this paper, we describe the first demonstration of an efficient Yb trapping
experiment in a conventional MOT by using the violet LD system. The performance characteristics of
our compact violet MOT, for example the trapped atom number of different isotopes, are found to be
similar to those of the previously reported Yb MOT by using a frequency-doubled Ti:Sapphire laser
\cite{Loftus2}, yet with a greatly reduced cost and complexity of the experiment. In addition, for
a future optical clock studies, we have successfully trapped 1.4 $\times 10^6$ fermionic $^{171}$Yb
atoms with a temperature of 0.7 mK, which is very close to the Doppler limit of 670 $\mu$K.

\begin{figure}
\includegraphics[height=6cm,width=5cm,angle=-90]{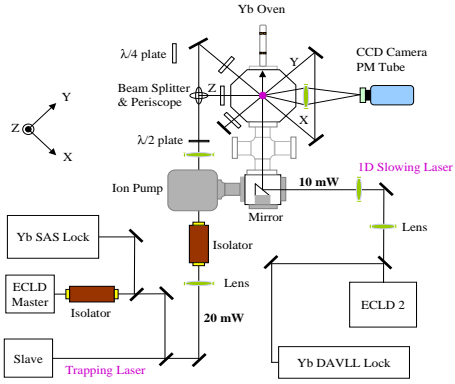}
\caption{Optical layout of the efficient trapping of a 1D-slowed Yb atomic beam with a violet LD
system, without a Zeeman slower.} \label{fig1}
\end{figure}

Figure~\ref{fig1} shows the layout of our experimental setup for the efficient trapping of Yb from
an atomic beam effusing from an oven, but without a Zeeman slower. The experiments were done inside
an ultra-high vacuum chamber consisting of a commercial stainless octagon. The octagon was pumped
by two 45 l/s ion pumps and maintained at a pressure of $4\times10^{-10}$ torr without Yb atoms and
$6\times 10^{-9}$ torr with the Yb atomic beam at the oven temperature of 390$^{\rm{o}}$C. The
trapping laser beam from the slave laser (described below) was divided into two beams by a beam
splitter for the horizontal axis (X and Y axis) and the vertical axis (Z axis). The horizontal beam
was retro-reflected in the X-Y plane, while the vertical beam was retro-reflected in the Z-axis, to
form a three-dimensional $\sigma^+$-$\sigma^-$ trapping geometry. Since the intensities $I$ of the
two trapping beams were only 5.0 mW/cm$^2$ just before the widows of the octagon, the saturation
parameter $s$ was very small, i.e., $s = I/I_s$ = 0.09, where $I_s =
2\pi\hbar\omega\Gamma/6\lambda^2 = 58$ mW/cm$^2$ is the saturation intensity, $\Gamma = 2\pi\times$
28 MHz (excited-state decay rate), $\omega = 2\pi$c/$\lambda$, c is the velocity of light, and
$\lambda$ is the wavelength of the trapping laser. Therefore, we attached the oven close to the
octagon (8 cm apart) to increase the atomic flux and used a 1D slowing laser to increase the number
of low velocity atoms below the capture velocity $v_c \approx 15$ m/s of the violet MOT
\cite{Oates}. At 390$^{\rm{o}}$C the fraction of Yb atoms under $v_c$ is estimated to be $3.5
\times 10^{-6}$. And by assuming the flux of Yb atoms from the oven is $10^{11}\sim 10^{12}$/s for
our trap geometry (nozzle diameter of 1.5 mm and 20 cm distance between the oven exit and the trap
center) \cite{Loftus}, we could roughly estimate the flux of Yb atoms slower than $v_c$ by using
the Maxwell-Boltzmann distribution, namely the loading rate of $10^4 \sim 10^5/$s without any 1D
slowing process. Therefore, the use of a 1D slowing laser is essential for the efficient trapping
of Yb atoms. In addition, to maintain a steep magnetic-field gradient and an ultra-high vacuum in
the octagon, we designed a compact anti-Helmholtz coil installed outside the octagon with coils
separated by only 4 cm. In the experiment, we found an optimal value of $\partial B/\partial z =
$45 G/cm at the coil current of 3.5 A and the trapping laser detuning of -0.5$\Gamma$.

To build a trapping laser at 398.9 nm, we developed a high-power violet LD system employing the
injection locking technique. Details of the master+slave violet laser system were previously
reported in \cite{Park}, and here we only briefly describe the system. The violet LDs used for the
master ECLD and the slave laser are high-power violet LDs with a 30 mW output power. The master
ECLD has a Littrow geometry in external feed-back and has an output power of more than 15 mW
\cite{Kim}. The frequency of the master ECLD could be locked to any of the dispersion slopes of the
Doppler-free absorption signals corresponding to the six stable Yb isotopes, i.e., isotopes 170,
171-174, and 176, obtained in the Yb saturation spectrometer (Yb SAS Lock). The Yb SAS consisted of
a Yb hollow-cathode lamp (HCL) with Ne buffer gas. However, we could not resolve the closely
positioned absorption lines corresponding to the two hyperfine transitions of the $^{173}$Yb
($^1$S$_0$(F=3/2) to $^1$P$_1$(F$^\prime$=3/2) and $^1$P$_1$(F$^\prime$=7/2)) and $^{172}$Yb
isotopes \cite{Park}. Since the Doppler-free width of 60 MHz is more than two times wider than the
natural linewidth of the $^1$P$_1$ excited state, we were easily able to tune the frequency of the
master laser to the optimal red detuning of -0.5~$\Gamma$ with an electronic servo system
\cite{Park}. We have estimated the frequency stability of the master ECLD from the measurement of
Allan variance of its error signal and obtained a frequency stability of 62 kHz at a 1 s average
time. Consequently, with a 0.6 mW injection power, the stabilized output of the master laser
increased coherently up to 25 mW with the slave laser. The beat-note linewidth at 160 MHz between
the master and slave lasers was measured to be below 1 Hz, indicating a good injection locking
quality with a stability of $2.5 \times 10^{-7}$ at a 1 s average time \cite{Park}.

A violet ECLD for the 1D slowing (ECLD2) was also constructed with the same high-power violet laser
diode. The frequency of the 1D slowing laser was stabilized to the broad Doppler-limited absorption
signal having a 2 GHz wide linewidth obtained with the modulation-free DAVLL (dichroic-atomic-vapor
laser lock) technique (Yb DAVLL Lock) \cite{Kim}. Due to its broad spectrum associated with the six
Yb isotopes, we were easily able to find the optimal detuning of the 1D slowing laser for a
specific isotope by changing electronically the reference voltage of the locking signal. A 10 mW
output of the ECLD2 was focused through a side window of the vacuum chamber as shown in
Fig.~\ref{fig1} to the exit of the oven collimator (not shown) \cite{Oates} to maximally overlap
the 1D slowing laser with the atomic beam at the center of the violet MOT, taking into account the
atomic beam divergence. With this focused 1D slowing laser, we achieved an increased acceleration
and power broadening simultaneously along the direction toward the exit of the oven, which resulted
in an increase of the total number of 1D slowed atoms in the trap site by as large as 35 times than
without the 1D slowing laser.

\begin{figure}
\includegraphics[height=6cm,width=5cm,angle=-90]{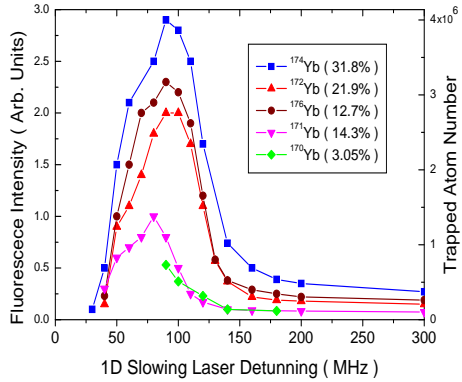}
\caption{Fluorescence intensity and trapped atom number as a function of detuning of the 1D slowing
laser for different Yb isotopes. The trapped atom numbers are calibrated from the spectral
sensitivity data of the PM tube and the imaging geometry in Fig.~\ref{fig1}. Inset shows the Yb
isotope atomic numbers and their natural abundances in parenthesis.} \label{fig2}
\end{figure}

Individual Yb isotopes could be sequentially loaded into the violet MOT by changing the locking
point of the trapping laser to correspond to a detuning of $-0.5\Gamma$ and slowly scanning the 1D
slowing laser frequency across the $^1$S$_0$-$^1$P$_1$ resonance. As a result, we were able to trap
almost all of the stable Yb isotopes. Figure~\ref{fig2} shows the fluorescence intensities for
different Yb isotopes as a function of the detuning of the 1D slowing laser from the frequency of
the trapping laser. We found that the optimal detunings of the 1D slowing laser were about -90 MHz
($\sim-3.2\Gamma$) for the bosonic isotopes, and -80 MHz ($\sim-2.9\Gamma$) for the fermionic
$^{171}$Yb isotope. The difference of the optimal detuning of the 1D slowing laser in the fermionic
isotope could be understood as resulting from optical pumping between the hyperfine sub-levels
present because of the non-zero nuclear spin. Note that the detuning of the 1D slowing laser in our
experiment is relatively close to the line center compared to the experiments using a Zeeman slower
\cite{Honda,Loftus}.

We measured the average cloud diameter of 2.5 mm (FWHM) by imaging the violet MOT onto the image
plane of the collection lens (see Fig.~\ref{fig1}) to measure the density of trapped atoms.
Fluorescence intensity from the MOT was easily measured with either a power-meter or a
photomultiplier (PM) tube located at the image plane of the trapped Yb cloud, enabling us to
calibrate and calculate the trapped atom numbers taking into account the solid angle of the imaging
lens and the scattering rate of the $^1$S$_0$-$^1$P$_1$ transition. The maximum trapped atom number
and density of the most abundant $^{174}$Yb isotope were about $4 \times 10^6$ and $1 \times
10^8$/cm$^3$, respectively. The vertical axis in the right-hand-side of Fig.~\ref{fig2} indicates
the calibrated trapped atom number. Table~\ref{table1} summarizes the maximum trapped atom numbers
and natural abundances for different Yb isotopes. Note that the trapped atom numbers listed in
Table~\ref{table1} are just a half smaller than those in \cite{Loftus2,Rapol}, where no 1D slowing
laser was used but a high-power (80$\sim$90 mW, $s =0.5\sim0.6$/axis) frequency-doubled Ti:Sapphire
laser for trapping was employed, demonstrating the high efficiency of the violet LD trapping of Yb
atoms with a 1D slowing laser. Furthermore, this result can also be explained with the earlier
observation of the power dependent loss mechanism in ref. \cite{Loftus}. The higher intensity of
the trapping laser increases the excited-state population $\rho_{22}$ and subsequently increases
the population loss in the Yb MOT through a weak branching via the $^3D_1$ and $^3D_2$ states to
the $^3P_2$ and $^3P_0$ metastable states resulting in the saturation of the total trapped atom
numbers at the low-level intensity ($s < 1$) of the trapping laser.

The fluorescence intensity ratios between the even isotopes (172, 174, and 176) were very similar
to the natural abundance ratios between them \cite{Honda,Loftus2}. However, for the fermionic
$^{171}$Yb isotope the trapping efficiency was 50\% smaller than those of the even isotopes
\cite{Honda,Loftus2}. In Fig.~\ref{fig2}, we omitted the fluorescence intensity corresponding to
the fermionic $^{173}$Yb isotope, since we could not observe a confined atomic cloud, rather we
were just able to detect the fluorescence from the optical molasses state. This fact can be
understood because the $^1P_1$ (F$^\prime$=7/2) hyperfine state of $^{173}$Yb lies too close to the
$^1P_1$ (F$^\prime$ =3/2) state (84 MHz apart), and thus optical pumping to the magnetic sub-levels
in the $^1S_0$ (F=5/2) ground state disturbs the cyclic cooling process \cite{Honda}. In addition,
in this level configuration the trapping beam is always blue-detuned from for the $^1P_1$
(F$^\prime$ =3/2) state, producing an opposite force relative to the trapping force by the $^1P_1$
(F$^\prime$=7/2) cycling transition due to the opposite sign of the Land\'{e} g-factors. This
problem could be solved by using the second-stage intercombination trapping transition at 555 nm
demonstrated in \cite{Honda}, where the excited hyperfine intervals of the three $^3P_1$ hyperfine
states of the $^{173}$Yb isotope are well separated ($>$ 1.5 GHz). The hyperfine splitting of the
excited state of $^{171}$Yb, however, is 319 MHz wide, therefore there is no optical pumping effect
in the $^{171}$Yb isotope. As shown in Fig.~\ref{fig2}, we have trapped 1.4$\times 10^6$ $^{171}$Yb
atoms with high efficiency (35\% of the most abundant $^{174}$Yb isotopes) which is slightly lower
than the result in \cite{Loftus2}, close to the natural abundance ratio of 45\% between them (see
Table~\ref{table1}). Consequently, we expect that in future studies of optical clocks based on the
fermionic $^{171}$Yb isotope, the violet LDs will be very useful for the the preparation of the
cold trapped $^{171}$Yb atomic samples.

\begin{figure}
\includegraphics[height=6cm,width=5cm,angle=-90]{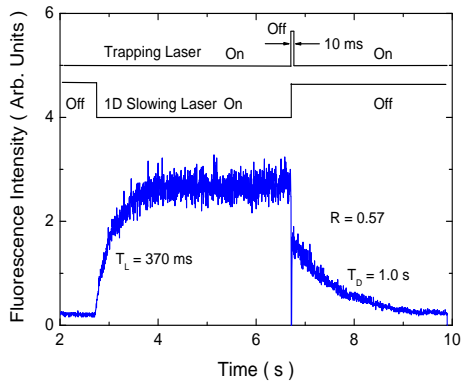}
\caption{Temperature measurement of the $^{174}$Yb isotope in the violet MOT by using a
release-and-recapture method along with the loading time T$_{\rm{L}}$ and decay time T$_{\rm{D}}$
measurements.} \label{fig3}
\end{figure}

To characterize the violet MOT further, we have measured the cloud temperature, the loading time
$T_L$, and the decay time $T_D$ of the MOT by using the well-known release-and-recapture method.
The bottom trace in Fig.~\ref{fig3} shows the typical experimental data for the loading time and
decay time measurements of the $^{174}$Yb isotope. Upper traces in Fig.~\ref{fig3} show the
chopping time sequences of the trapping and 1D slowing lasers for two mechanical shutters installed
in front of the slave laser and ECLD2 in Fig.~\ref{fig1}. After loading the MOT with the trapping
laser only, we suddenly turned on the 1D slowing laser at T = 2.7 s, as shown in Fig.~\ref{fig3},
to see the increase of fluorescence intensity with the help of the 1D slowing laser. From the
exponential fit to the data, we could obtain the loading time of $T_L$ = 370 ms. As one can see
from Fig.~\ref{fig3}, the use of the 1D slowing laser increased the trapped atom number by about 35
times more than that obtained with the trapping laser only. After the steady state was reached, we
abruptly turned off both lasers at T = 6.7 s to let the atomic cloud freely expand, and 10 ms later
we turned on the trapping laser again, thereby re-capturing the atoms by the trapping laser only.
Then, the fluorescence intensity decays with the same decay rate of the MOT until the steady-state
reached. From the exponential fit to this decay curve, we obtained the decay time of $T_D$ = 1.0 s.
We repeated the measurement for the different Yb isotopes and listed the measurement results in
Table~\ref{table1}. The difference between the loading rate and decay rate in Fig.~\ref{fig3} and
Table~\ref{table1} is caused by the fact that the detuning of our 1D slowing laser is only
3$\Gamma$ from the line center and there is a 1D radiation pressure which increases the loss rate
of the violet MOT when the 1D slowing laser is turned on.

Since we know the size of the trapping beam (1 cm), we can estimate the cloud temperature by
measuring the fluorescence intensity ratio $R$ before and after the 10 ms interval at T = 6.7 s in
Fig.~\ref{fig3}. If we assume a three-dimensional isotropic Boltzmann velocity distribution at the
given temperature T, and calculate the time evolution of the distribution after 10 ms taking into
account the increased cloud size, then it is straightforward to obtain the ratio between the number
of atoms within the size of the trapping laser beam to the total number of initially trapped atoms.
This ratio should be directly proportional to the fluorescence intensity ratio depending on the
initial cloud temperature as measured in Fig.~\ref{fig3}, i.e., R = 0.57 for $^{174}$Yb. By
adjusting the initial temperature of the trapped atoms in a fitting program until the calculated
ratio of the atomic numbers agrees with the measured ratio R of the fluorescence intensities, we
can determine the temperature of the trapped atoms. Table~\ref{table1} also shows the measured
fluorescence intensity ratio (recapture ratio) R and the cloud temperature of the different Yb
isotopes. We found that the cloud temperatures for most Yb isotopes were slightly above the Doppler
limit of 670 $\mu$K as can be seen in Table~\ref{table1}. The relatively low temperature of the
trapped Yb atoms observed resulted from the low saturation parameter (s = 0.09) of the violet LD
trapping laser. However, the focused 1D slowing laser increased the total number of trapped atoms,
comparable to the results in \cite{Honda} and \cite{Loftus2}.

In summary, we have efficiently cooled and trapped the rare-earth Yb atoms in a compact violet
magneto-optical trap employing a high-power frequency-stabilized violet LD system without a Zeeman
slower. The overall characteristics of our compact violet Yb MOT are found to be similar to those
of the previously reported Yb MOTs using a frequency-doubled Ti:Sapphire laser and a Zeeman slower,
yet with greatly reduced cost and complexity. Importantly, for future optical clock studies, we
have successfully trapped 1.4 $\times 10^6$ fermionic $^{171}$Yb atoms with a temperature of 0.7
mK, which is very close to the Doppler limit of 670 $\mu$K. We are now trying to employ a far-off
resonant dipole trap, following the Katori's proposal \cite{Katori}, to realize a Lamb-Dicke
confinement of neutral $^{171}$Yb fermionic atoms for a new Yb optical clock.

This research was supported by the Creative Research Initiatives Program of the Ministry of Science
and Technology of Korea. The authors thank J. Y. Yeom and S. Pulkin for help and support during the
course of this research and C. Sukenik for critical reading of the manuscript.


\begin{table}[h]
\caption{Natural abundance (NA), maximum trapped atom number (TN), trapped atomic temperature (CT),
loading time (LT), decay time (DT), and re-capture ratio (RR) for different Yb isotopes in the
violet MOT.}\label{table1}\begin{center}
\begin{tabular}{c|c|c|c|c|c|c}\hline
Isotope & NA (\%)&TN & CT (mK) & LT (s) & DT (s) &  RR \\
\hline
170 & 3.1&$7.0\times10^5$ & 1.0 &0.8 &0.5&0.4 \\
171 & 14.3&$1.4\times10^6$ & 0.7 &0.7 &0.6&0.6 \\
172 & 21.9&$3.2\times10^6$ & 0.7 &0.5 &0.9&0.6 \\
174 & 31.8&$4.9\times10^6$ & 0.7 &0.4 &1.0&0.6 \\
176 & 12.7&$2.8\times10^6$ & 0.7 &1.0 &0.9&0.6 \\ \hline
\end{tabular}
\end{center}
\end{table}

\end{document}